\documentclass[%
 reprint,
showpacs,preprintnumbers,
 amsmath,amssymb,
 aps,
pra,
floatfix,
]{revtex4-1}

\usepackage{graphicx}
\usepackage{subfig}
\usepackage{dcolumn}
\usepackage{bm}
\usepackage{enumerate}

\begin{document}

\preprint{APS/123-QED}

\title{Quantum Random Number Generator based on Tunneling effects in Si Diode}

\author{Haihan Zhou}
 \email{zhouhh15@mails.tsinghua.edu.cn}
\author{Weixing Zhang}%
\affiliation{%
 Physics Department, Tsinghua University, Beijing, PRC}
\author{Junlin Li}
 \email{center@mail.tsinghua.edu.cn}
\affiliation{
  Physics Department, Tsinghua University, Beijing, PRC
}%
\author{Gui-Lu Long}
 \email{gllong@tsinghua.edu.cn}
\affiliation{
  Physics Department, Tsinghua University, Beijing, PRC
}%

\date{\today}

\begin{abstract}

Previously, we built up a set of photon-free quantum random number generator(QRNG) with InGaAs single photon avalanche diodes. We exploited the stochastic property of quantum tunneling effect. Here, we utilized tunneling signals in Si diodes to implement quantum random number generator. In our experiment, instead of applying periodic pulses between the diode as we did in the InGaAs QRNG, we applied fixed voltage and detect time intervals between adjacent tunneling signals, as random source. This Si QRNG has a high performance in the randomness of its raw data and almost post-processing-free. Final data rate in our experiment is 6.98MB/s and could reach 23MB/s if the temperature-control system is ameliorated.      

\end{abstract}

\maketitle


\section{\label{sec:level1}Introduction}

Randomness is one of the most widely used data property in physics, mathematics and computer science, so does their derivative subjects. Specifically, in many fields like  machine learning\cite{robert2014machine}, cryptography\cite{bennett1991experimental}, quantum computation\cite{paz2003randomness} and quantum information\cite{deng2004secure}, true randomness is indispensable. Numerous studies has been conducted hitherto on the randomness examination\cite{soto2000randomness} and random number generation\cite{park1988random}. 

Quantum random number generators exploited many theoretical uncertainty, assured by the basic principles in quantum mechanics\cite{erber1985randomness}, to establish eligible random number generation systems. In year 2000, 'Path-choice' of single polarized photon after passing a polarized beam splitter was applied in the design of quantum random number generator\cite{jennewein2000a}. Later, various schemes rose up and different systems are all utilized as random source, such as photon arrival time\cite{wayne2009photon}\cite{ma2005random}, phase fluctuation of vacuum state\cite{xu2012ultrafast}, quantum phase noise\cite{qi2010high-speed}. The generation speed of quantum different schemes varies too, most discrete cases could not reach a higher speed than 100MB/s, while continuous schemes reached GB/s\cite{article}. Later, concepts of Bell inequality was applied to QRNG designs to assure the true randomness. In 2015, self-testing quantum random number generator was designed by Lunghi and Bowles. They designed a discrete quantum random number generator that can continuously measure its output entropy via estimation of 'dimension witness'\cite{lunghi2015self-testing}. Following this work, Prof.Ma proposed a concept of semi-self-testing QRNG and designed experiments in single photon system\cite{cao2016source-independent}. 

In our previous study\cite{2017arXiv171101752Z}, we utilized the randomness of quantum tunneling effects in InGaAs diodes in the design of quantum random number generator. Based on the collisional ionization theory, we can estimate the tunneling probability in each part of the SACGM InGaAs/InP SPAD\cite{tarof1990planar}. We managed a 15MB/s quantum random number generator. In this paper, we still took the advantage of quantum tunneling effects. Si SPAD had a simpler structure as there is no band gap difference between the  absorption layer and multiplication layer\cite{cova2004evolution}. Furthermore, we simplified the experiment set-up. Instead of applying a periodic bias voltage on the  InGaAs diode, we fixed the bias voltage on Si diode and detect the time interval of adjacent tunneling signal. After data pre-selection, the min-entropy could reach 9.8 each 10 bits. The output data passed NIST\cite{Rukhin2010A} and Diehard tests\cite{Marsaglia1995The} after randomness extraction\cite{L2007TestU01}.

\section{Collisional ionization in Si SPAD}
Theory model on free charge carrier in semi-conductor diodes has been studied since 1950s\cite{Anthony1958Semi}\cite{11961Fluctuation}\cite{Noyce1961Semiconductor}. Collisional ionization is the most typical theory\cite{Mcintyre1966Multiplication}. McIntyre proposed  this model in 1960s. Free carrier like electrons or holes would make collision during the propagation in semi-conductors. These collisions would excite more carriers. Generally, initiation of these free carriers would be generated under many circumstances, such as  photon absorption, thermal excitation, tunneling effects and after-pulse effects. Here, in our scheme, we restrained other effects to assure the tunneling effect was the major cause of dark signals. During our experiments, the optical input port was closed to prevent photon absorption. Also, we utilized a semi-conductor cooling system to maintain the working temperature of Si SPAD. Specifically, the system was cooled to $-20$ $^{\circ}$C in experiment. Furthermore, we applied active-quenching system to decrease the after-pulse effect. We have set the hold-off time after each tunneling signal to $17ns$. Yet, we still could not eradicate the noise brought by it without pre-selection.  

Most Si SPADs are $p$-$i$-$n$ diodes, as the absorption layer has the same band width with multiplication layer. This simple structure made the analysis of tunneling probability much easier than InGaAs/InP SPAD. 

\begin{figure}
\centering
\includegraphics[width = 0.48\textwidth]{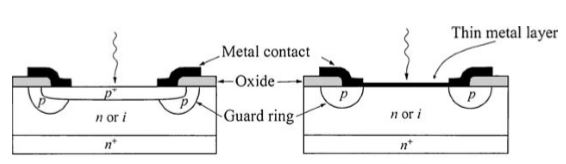}
\label{pinstructure}
\caption{Structure of p-i-n diode}
\end{figure}

According to McIntyre, electrons and holes had two certain collision probability in a unit length in semiconductor diodes $\alpha_e$ and $\alpha_h$. Then we can calculate the mean collision times $M(x)$ at a specific position inside the intrinsic layer of a $p$-$i$-$n$ diode:

\begin{align}
M(x) = \frac{exp(-\int_{x}^{L} \alpha_e - \alpha_h dx')}{1 - \int_{0}^{L} \alpha_e exp(-\int_{x'}^{L} \alpha_e - \alpha_h dx'' )dx'}
\end{align}
Here, $0$ and $L$ represent the interface of intrinsic layer with $p$layer and $n$layer. This equation was first derived by McIntyre.\cite{Mcintyre1972The}

The collision ionization theory pointed out the generation of new free carriers inside a diode is probabilistic, and gave us a  
mean estimation of the average collisions. Another probabilistic process is the avalanche. The probability of avalanche caused by a single carrier could also calculated under this model:

\begin{align}
P_p(x) &= \frac{P_p(0)f(x)}{P_p(0)f(x)+1-P_p(0)} \\
P_p(0) &= 1- exp(-\int_{0}^{L} \alpha_h PA(x') dx') \\
f(x) &= exp(-\int_{0}^{x} \alpha_e - \alpha_h dx')
\end{align}

Here, $L$ is the length of depletion layer. Since the probability of avalanche was calculated. One of the most vital parameter, dark count rate(DCR), can now be calculated. The majority of DCR was consisted of thermal noise and tunneling-generation noise. 

Generally, the thermal noise is determined by intrinsic property of materials and temperature:

\begin{align}
G_{Ti}  = \frac{n_i}{\tau_i}
\end{align}

$i$ denotes the layer; $n_i$ represents the intrinsic carrier in  layer $i$ and $\tau_i$ stands for the life time of free carriers in layer $i$. $n_i$ and $\tau_i$ are both under the influence of temperature. 

Tunneling-generation noise is more complicated than thermal noise. It includes two part signals: Band-Band-Tunneling signals\cite{Schenk1993Rigorous} and Trap-assisted-tunneling signals\cite{M1990Model}. This means free carriers could tunnel directly through band gap or take a two-step tunneling with a trap state as its medium. Thus the number of free carriers in diode could be considered as follows\cite{Donnelly2006Design}:

\begin{align}
N_{i} &= \frac{J_{BBT} + J_{TAT}}{q} \\
J_{BBT} &= \sqrt{\frac{2m_r}{E_g}}\frac{q^2 F^2}{4\pi^3 h^2} exp(-\frac{\pi \sqrt{m_r E_g^3}}{2\sqrt{2}q \hbar F}) \\
J_{TAT} &= \frac{\sqrt{\frac{2m_r}{E_g}}\frac{q^2 F^2}{4\pi^3 h^2} N_{trap} exp(-\frac{\pi \sqrt{m_{lh} E_{B1}^3} + \pi \sqrt{m_c E_{B2}^3}}{2\sqrt{2}q \hbar F})}{N_\upsilon exp(-\frac{\pi \sqrt{m_{lh} E_{B1}^3}}{2\sqrt{2}q \hbar F})+ N_c exp(-\frac{\pi \sqrt{m_{c} E_{B2}^3}}{2\sqrt{2}q \hbar F})}
\end{align}

$q$ is the carrier charge; $m_r$ is the effective mass; $E_g$ is the bandgap; $F$ is the electric field, which is a function of position. $N_{trap}$, $N_\upsilon$ and $N_c$ are the density of traps in a unit volume, density of light-hole states in valance and conduction band, respectively. $m_{lh}$ and $m_c$ are the effective mass of light-hole and conduction band. $E_{B1}$ and $E_{B2}$ are the energy gaps from valance band to trap and from trap to conduction band. 

The DCR of diode could be expressed with the integration of $G_{Ti}$ and $N_{i}$ through the whole absorption layer and depletion layer.
\begin{align}
DCR = \int (G_{Tdep} + N_{dep})P_p(x')\ dV_{dep} \notag \\+ P_p(0) \int G_{Tab} + N_{ab}\ dV_{ab}
\end{align}
 
DCR gave us the number of dark signals from Si diode in a unit time. Furthermore, accounting on the practical situation, we assumed the $G_{Ti}$ to be far smaller than $N_{i}$, $\alpha_e$ and $\alpha_h$ to be constants under a fixed bias voltage.

As in our experiment, what we focused on is the application of Si SPAD in quantum random number generation. Thus, we did not lay our points on the doping of each layer. More specifically, we chose a Si SPAD from a Si single photon detector produced by a company in Shanghai and did not change the inner structure of it in the following experiments. 

\section{Scheme}
Previous studies on collisional ionization demonstrated the process of signal generation inside a diode. Among dark counts of a diode, tunneling signal could be a major part in low-tempreture case. When we apply proper bias voltage on Si SPAD, electrons inside would tunnel through the valance band and conduct band with a certain probability and trigger avalanche signals. Generally, tunneling counts consisted of two parts, Band-Band-Tunneling part and Trap-assisted-Tunneling part. In our previous study, we manipulated tunneling probability to $0.5$ via adjustment of bias voltage and collected random sequence. This design made the data collection and data analysis quite direct. Yet, it requires  an extreme stable trigger clock and synchronization system. On the other hand, tunneling probability in Si diode could be much lower than in InGaAs/InP diode. Therefore, we took the time interval of adjacent tunneling signals as random source, we could use a much stable DC power to provide the bias voltage.      

Basic thoughts of our experiment is under that a certain bias voltage on Si SPAD would cause tunneling counts with certain probability. This tunneling signal could get amplified inside the Si diode and then get detected. The exact time of the tunneling signals' arrival is unpredictable. Random numbers are generated from these successive undetermined signals. Detailed scheme can be expressed as follows.

\begin{enumerate}[Step 1]
\item We first cool the Si SPAD down to -20$^{\circ}$C, and apply a reverse bias voltage on it. 

\item Adjust the power to an 'efficient' voltage. The detector would receive signals.

\item Connect the signal collection module with a standard clock. We would output the number of periods between adjacent tunneling signals as random signals.

\item Connect the data collection module to a PC, and run the data collection software to record the number of tunneling in a second.

\item After the generation speed stabilized, start data collection. 

\item Pre-select the data, and conduct randomness extraction.

\item Check the Randomness of final data.

\end{enumerate}

\begin{figure}
\centering
\includegraphics[width= 0.48 \textwidth]{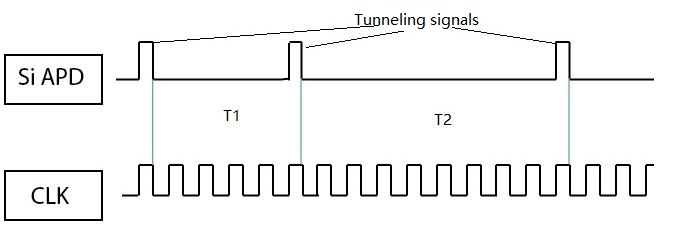}
\label{Sitime}
\caption{Simple diagram of how we record the time intervals}

\end{figure}
As we measure the time interval of adjacent tunneling signals by a standard clock, whose frequency is 500MHz, we assume that during $1$ period of this clock, the probability of electron tunneling is $p$ and very small. $p$ here is a function of several factors.

\begin{align}
p = p(U,T,...)
\end{align}

As we fixed our the output of the DC power, we can further say:

\begin{align}
p(U,T,...) = p_0
\end{align}

If each tunneling signal has been independent in the system, we can directly come to the probability that time interval $t = n\times T$ between adjacent signals under a fixed voltage$U_0$:

\begin{align}
P(n) & = (1-p_0)^n p_0 \notag \\
& = p_0 (1-p_0)^{\frac{n}{p_0} p_0} \notag \\
& = p_0 e^{-np_0} 
\end{align}

Here $T$ refers to the period of standard clock, so the probability obeys an exponential distribution. However, due to the after pulse effect of Si diode\cite{Dalla2012Afterpulse}, every tunneling signal would leave electron-vacancy pairs in the diode, thus enhance the tunneling probability in next periods. Furthermore, this effect should be short-range in time, that is, for a sufficient long time after former tunneling signal, this effect can be omitted. Under this estimation, we can rewrite the probability as:

\begin{align}
P(n) = (p_0+p_e(t_n))\prod\limits_{i = 1}^{i = n-1}(1-p_0-p_e(t_i))
\label{P}
\end{align}

$p(t_i)$ refers to the additional probability induced by a tunneling signal after $i$ clock periods. We would conduct our data pre-selection based on equation above later section. 

Before we make detailed analysis of $p_e(t_i)$, we first collected the 60MB output of Si SPAD under $98V$, and drew a dark counts-time diagram with theoretical results versus experimental results as shown:
\begin{figure}
\centering
\includegraphics[width=0.48 \textwidth]{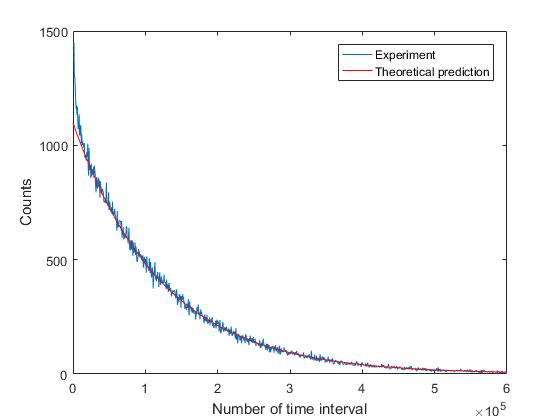}
\caption{Theoretical prediction v.s. Experimental results under 98V, 60MB data. }
\label{98}
\end{figure} 
Despite of differences when $n$ nears $5\times 10^4$, they are highly fitted. Thus, we can make a preliminary process on this data to transform them to an even distribution. We divided the area covered by the theoretical line into 1024 equal fractions. Then encode all the experimental data in each faction with the binary form of the faction's serial number. The basic principle is like:
\begin{figure}
\centering
\includegraphics[width=0.48 \textwidth]{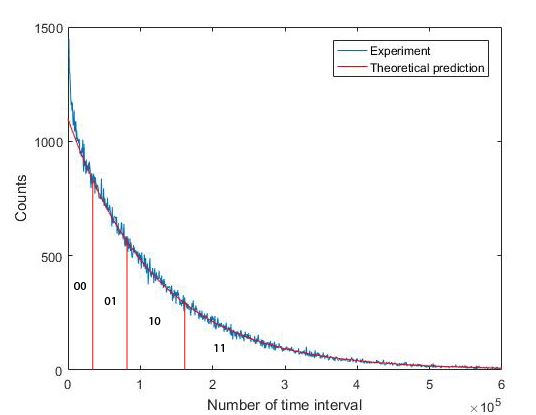}
\caption{Simple concept of preliminary encoding on output data.}
\label{divide}
\end{figure}
We applied this concept on a part of 1.5GB data collected under 100V, and got:
\begin{figure}
\centering
\includegraphics[width=0.48 \textwidth]{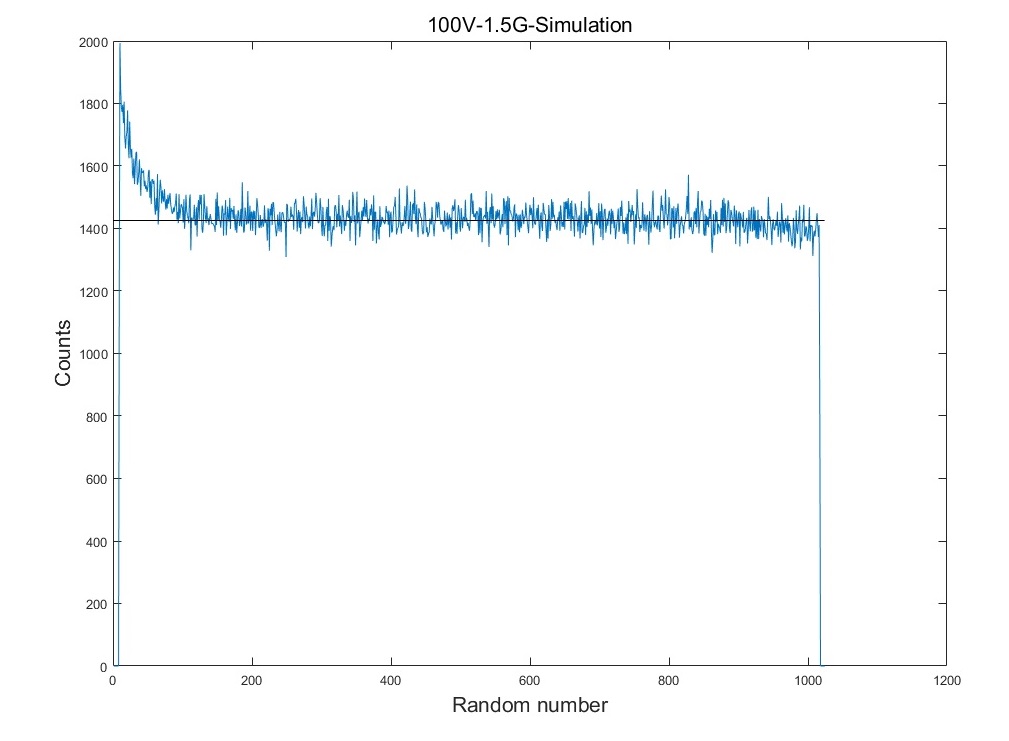}
\caption{Results of data after encoding}
\label{ResultsAfter}
\end{figure}

As we can see from \ref{ResultsAfter}, counts of random numbers near 0 is went beyond the theoretical prediction. Thus, we have to take $p_e{t_i}$ into consideration to eliminate this difference. This would be illustrated in the next section.

\section{Set-up, Pre-selection and Post-processing}
\subsection{Experimental Set-up}
In this section, we will illustrate structure of our Si QRNG. As stated in our scheme, we mainly have 4 parts in our Si quantum random number generator: The Si SPAD part, the power and cooling system, the active-quenching system, the pre-selection and post-processing system. 

We exploited a semi-conductor cooling system to maintain the working temperature of Si SPAD was $-20$ $^{\circ}$C, getting rid of the thermal noise. This system was tested to be stable after a 4-hour continuous running.

Except from the thermal noise, another main adversary of our system is the after-pulsing effects. As we illustrated before, free carriers could generate more of them with the process of collisional ionization. Unfortunately, each success of tunneling effect means there would be some remained free carriers, captured by the inner defects of diodes. These captured would get freed as those created them, only with a delay. This delay contributed to the increase of successive dark counts' probability after a real tunneling signal. 

\subsection{Pre-selection and Post-processing}
The magnitude of after-pulse is associated with several factors. The probability of after-pulse can be expressed as:
\begin{align}
P_{a} \propto C T exp(-\frac{t}{\tau})
\end{align}
$C$ is the sum of Si SPAD's effective capacitance and parasitic capacitance of the whole circuit. $T$ refers to the duration of each avalanche. $t$ is the 'hold-off time', here $17ns$. $\tau$ is the lifetime of free carriers. Once our system was set, $P_a$would be a exponential to $t$. We can further simplify it to:
\begin{align}
P_a(t) = A exp(-Bt)
\end{align}

In light with such ubiquitous noise in most diodes, people designed different systems to get rid of it. Basically, there are three main methods: passive-quenching, active-quenching and gate-pulse-quenching\cite{Lacaita1996Avalanche}.
  
\begin{figure}[!htb]  
  \centering  
  \subfloat[Active-quenching]{\includegraphics[width=0.25\textwidth]{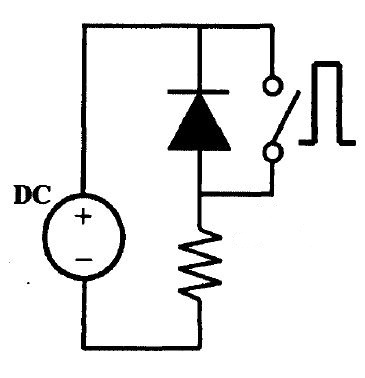}} \   
  \subfloat[Passive-quenching]{\includegraphics[width=0.2\textwidth]{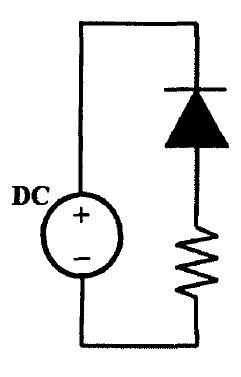}} \\  
  \subfloat[Gate-quenching]{\includegraphics[width=0.25\textwidth]{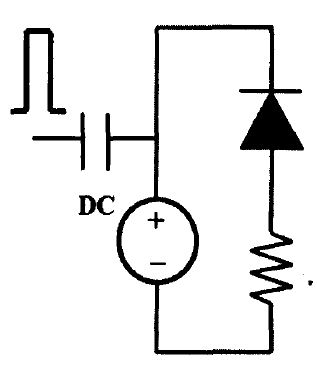}}   

  \caption{Three different ways of eliminating after-pulse.}    
    \vspace{0.2in}  
\end{figure} 
Each of these methods has advantages and drawbacks. In our system, we chose the active-quenching circuit. Namely, after an avalanche, the voltage on SPAD would be lowered to $0V$ for $17ns$ to decrease the after-pulse. This circuit protect Si SPAD from continuous Geiger mode and have a fast corresponding time. However, the on-off of the switch would cause sharp noises. And it would take a more complex circuit to reduce such noise, which would not be elaborated here. 

We can now use $P_a(t)$ to substitute $p_e(t_i)$ in \ref{P}. For further analysis, we can take logarithm of \ref{P}:
\begin{align}
ln(P_r(n)) = ln(p_0 + P_a(t)) + \sum_{i=1}^{i=n-1}ln(1-p_0-P_a(t_i))
\label{lnP}
\end{align}

As the $n$ here means the number of periods counted by standard clock, also the periods is quite short, $2ns$. We can take rewrite \ref{lnP} as with integration on $t$:
\begin{align}
ln(P_r(t)) = ln(p_0 + P_a(t)) + \int_{0}^{t} ln(1-p_0-P_a(t'))\  dt'
\end{align}
Furthermore, we have $P_a(t) \ll p_0 \ll 1 $. 
\begin{align}
ln(P_r(t)) = ln(p_0 + P_a(t)) - \int_{0}^{t} (p_0 + P_a(t')) dt' \\ P_r(t) = (p_0 + P_a(t))exp(-\int_{0}^{t} (p_0 + P_a(t'))\ dt')
\end{align}

In order to attest our theory, we calculate the quotient of $P_r(t)$ on $P(t)$.

\begin{align}
\frac{P_r(t)}{P(t)} = \frac{p_0+P_a(t)}{p_0} exp(-\int_{0}^{t} P_a(t')\ dt')
\end{align}

Then calculate the logarithm of this quotient:
\begin{align}
ln(\frac{P_r(t)}{P(t)}) = ln(\frac{p_0+P_a(t)}{p_0}) - \int_{0}^{t} P_a(t')\ dt'
\label{logofquo}
\end{align}
Consider the fact that $P_a(t) \ll p_0$, and bring the concrete expression into \ref{logofquo}:
\begin{align}
ln(\frac{P_r(t)}{P(t)}) &= \frac{Aexp(-Bt)}{p_0} + \frac{A}{B} exp(-Bt) \notag \\
&= \frac{A(p_0+B)}{p_0 B} exp(-Bt)
\end{align} 
So under the assumption that $P_a(t) \ll p_0 \ll 1$, this logarithm of quotient should be a exponential of $t$. Then we calculate this value for output data under 94V, 98V, 100V, and 102V, respectively.

\begin{figure}[!htb]  
  \centering  
  \subfloat[Log of quotient under 94V]{\includegraphics[width=0.23\textwidth]{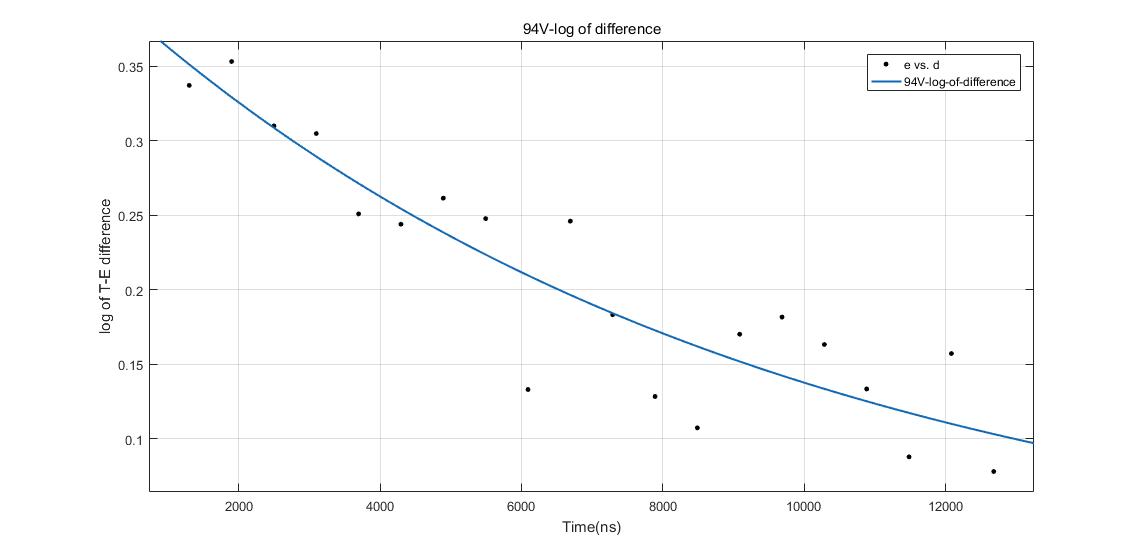}} \   
  \subfloat[Log of quotient under 98V]{\includegraphics[width=0.23\textwidth]{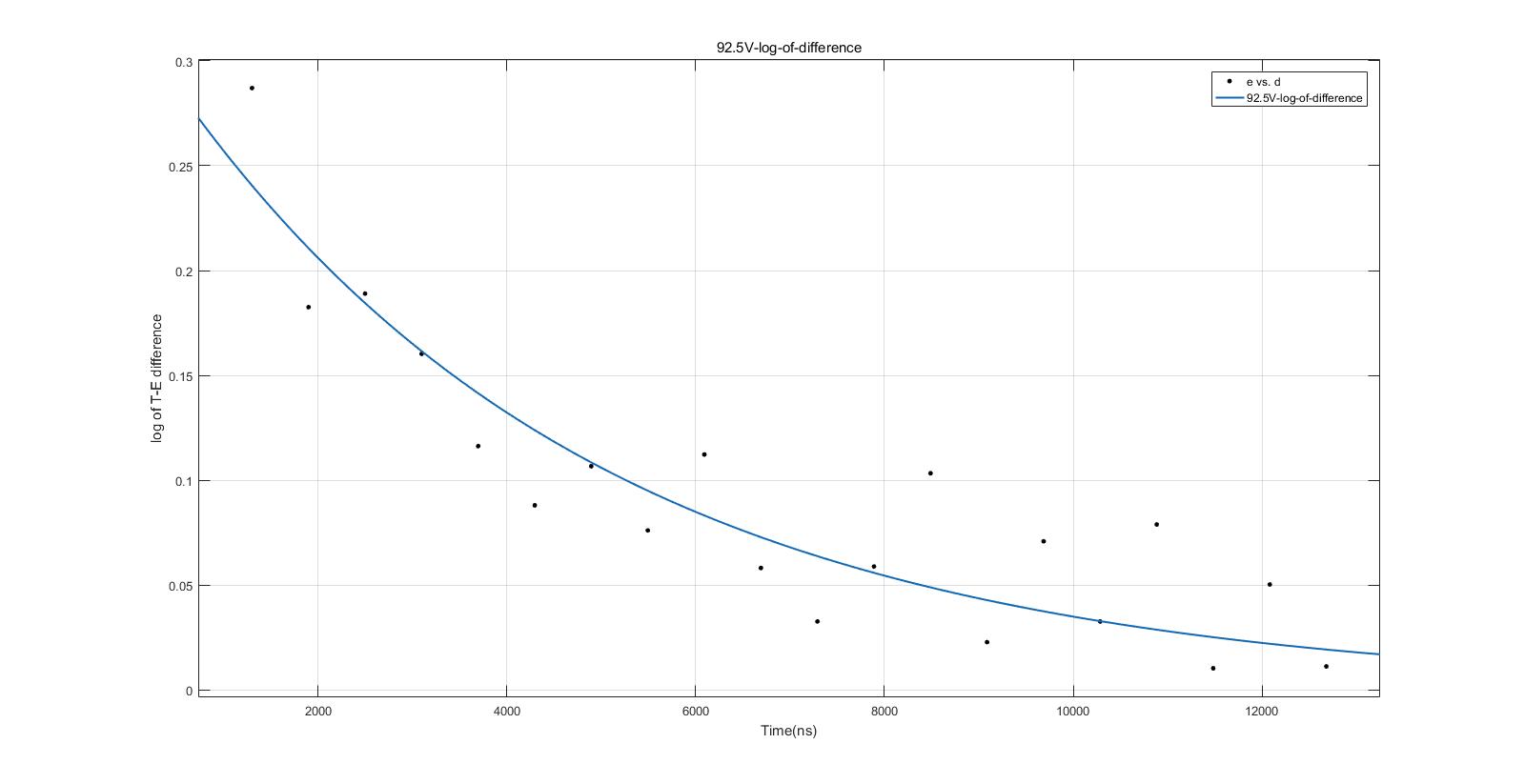}} \\  
  \subfloat[Log of quotient under 100V]{\includegraphics[width=0.23\textwidth]{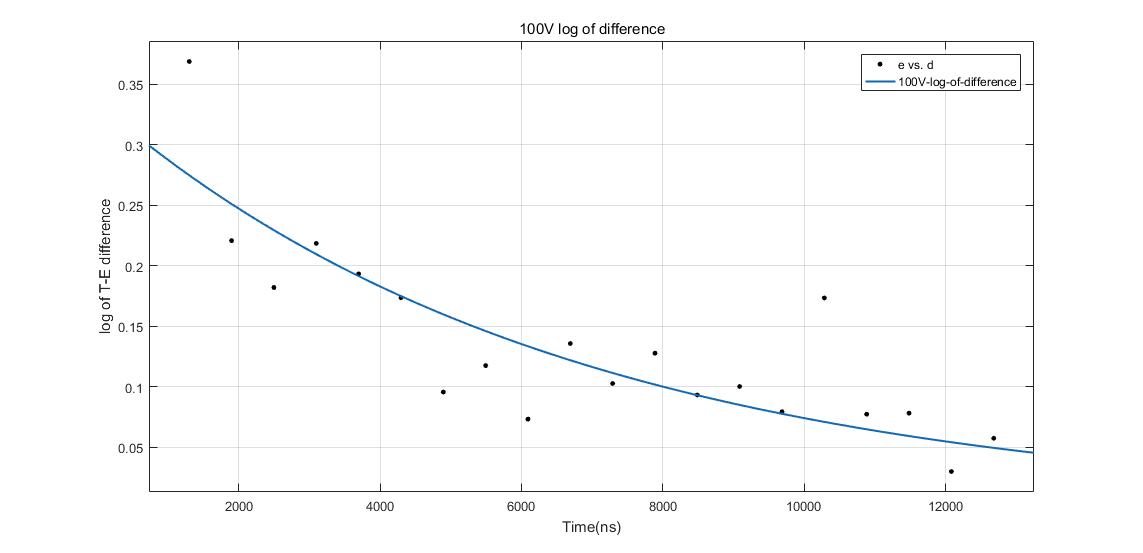}}   
\subfloat[Log of quotient under 102V]{\includegraphics[width=0.23\textwidth]{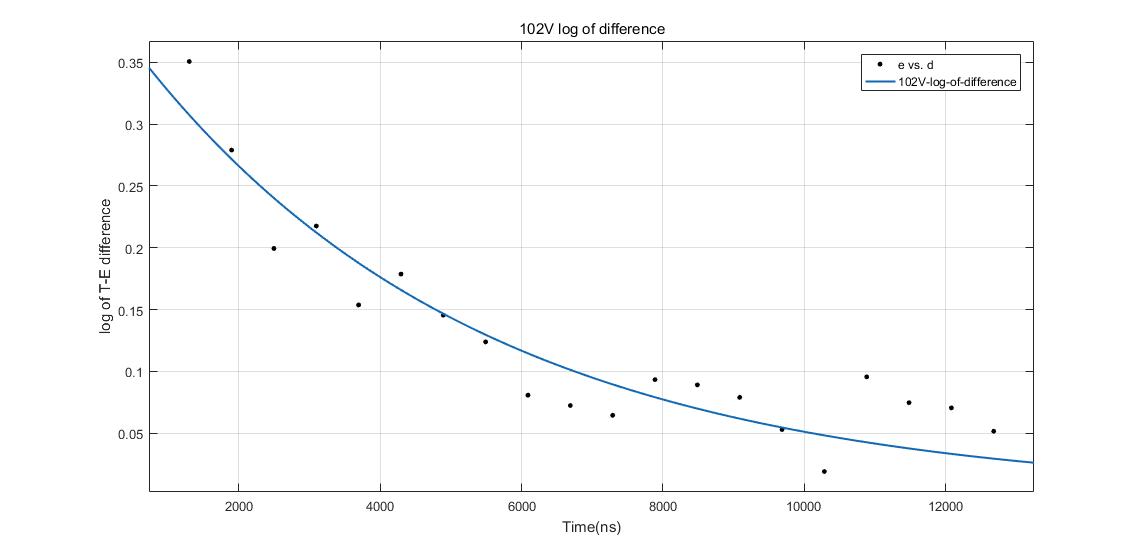}}   
  \caption{Theoretical prediction versus experimental results on log of quotient under different voltage.}    
    \vspace{0.2in}  
\end{figure} 
We can find out that the experimental results fits the theoretical prediction under these different bias voltages. Thus, We can now correct the deviation mainly caused by after-pulse effects. As we take this noise model into consideration, we can cut the number of counts at each random number by specific proportion. We call this pre-selection before post-processing. We, again, pre-select the 1.5GB data under 100V, and then take a part of it to show the distribution, and the result can be shown as:
\begin{figure}
\centering
\includegraphics[width=0.5 \textwidth]{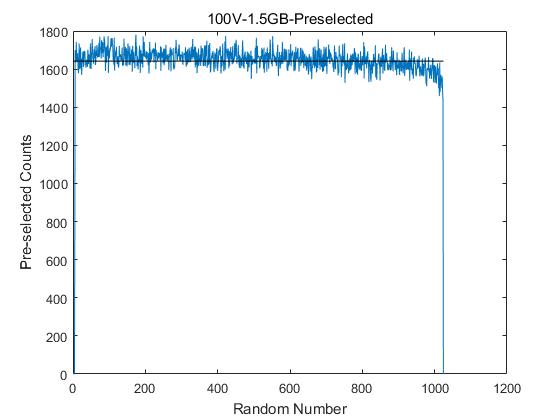}
\label{Preselected}
\caption{Distribution of pre-selected random number}
\end{figure}

One of the most vital parameter in post-processing -- the min-entropy\cite{Wiggins1980Minimum} is also optimized after the pre-selection. Randomness extractors based on min-entropy have been widely used in previous work on QRNG\cite{Chevassut2005Key}\cite{Kiltz2009A}\cite{Dodis2004Improved}. In our scheme, we used Toeplitz-hashing extractor for randomness extraction\cite{Ma2012Postprocessing}. For a sequence consisted by random variable $X$, the definition of min-entropy is :
\begin{align}
H_m = -log_2 (max{X})
\end{align}
Min-entropy represents the disorder of the sequence, and the more chaotic the sequence is, the more information $X$ could take. Extractors are supposed to distill the randomness in $X$ and omit the rest. They often work in following pattern:
\begin{align}
\lbrace 0,1\rbrace ^ m \bigotimes Seed   \xrightarrow{extract}\lbrace 0,1\rbrace ^ n
\end{align}
Toeplitz-hashing extractor and Trevisan extractor are both widely used extractors. In the application of Toeplitz-hashing extractor, we have to build a Toeplitz matrix based on the min-entropy, then calculate the product of output data and this matrix to get the final data. 

In our experiment, we calculated the min-entropy each 10-bit of our output data. After pre-selection, min-entropy was 9.79. This means we can preserve near $98 \%$ data after the post-processing, which indicated our system is rather efficient. We would test the final data with Diehard and NIST in next section.

\section{Randomness Test}
After the post-processing, we finally got a 1.5GB data with 100V bias voltage. The next step is randomness test, which check the eligibility of our system. There have been several kinds of randomness tests. For instance, NIST Test suite\cite{Rukhin2010A}, Diehard test\cite{Marsaglia1995The}, TestU01\cite{L2007TestU01}. Each of them contained subsections of more explicit statistical tests, verifying the randomness of input data from different aspects. 

Here, we used NIST test and Diehard test to check the randomness of final data. 
\begin{figure}
\centering
\includegraphics[width=0.48 \textwidth]{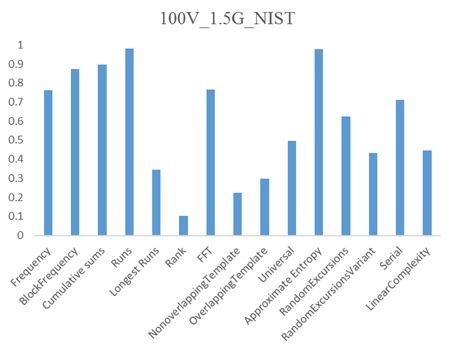}
\label{NISTtest}
\caption{Result of NIST test of 1.5GB data under 100V}
\end{figure}

\begin{figure}
\centering
\includegraphics[width=0.48 \textwidth]{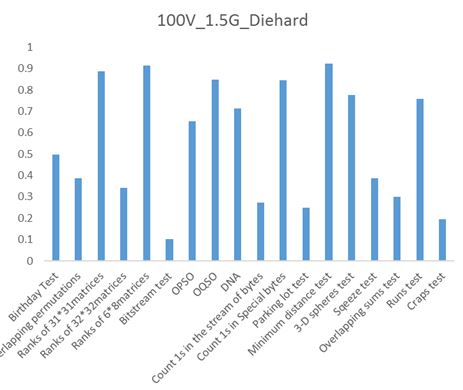}
\label{Diehardtest}
\caption{Result of Diehard test of 1.5GB data under 100V}
\end{figure}

The criteria of passing NIST was the p-value of each test lies between 0.01 and 0.99. Our data met this criteria with passing rate $98.2 \%$. The criteria of passing Diehard test was the p-value lies between 0.000001 and 0.999999. Concrete p-values were listed in the Appendix. 

As shown, our final data passed this two tests. Another advantage is that all the data needed in our data encoding, pre-selection and post-processing is actually the number of DCR and the output data itself. So, the software needed in data collection is quite simple.

\begin{figure}
\centering
\includegraphics[width=0.48 \textwidth]{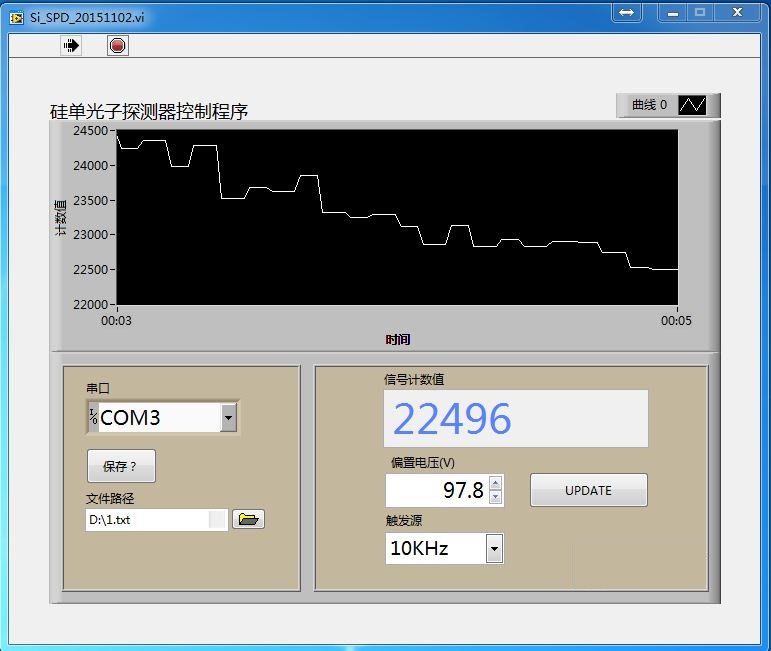}
\label{Software}
\caption{Software designed for data collection. It records the time intervals and the DCR per second.}
\end{figure} 

So far, we have implemented a Si QRNG based on the tunneling effects with output speed 6.98MB/s. This system, with quite simple and cheap set-up, can work stably for more than 4 hours.

\section{Conclusion}

This paper introduced our work on practical design of Si QRNG based on tunneling effects. Our system is quite simple and cheap in comparison with many other QRNG with photon source. The performance of our system could also be enhanced by improving hardware of each module, which would be studied in our future work.

There are some other aspects future work might focus on. Despite the output speed of our system could be enhanced, it could not exceed most continuous QRNG. This means our system could not serve many high-speed circumstances. It is not the problem of tunneling effects but the Si diodes system. Other system should be studied to proceed higher generation speed. 

Another problem is that the deviation caused by after-pulse noise has been illustrated but not efficiently solved. The reverse procedure of cutting this deviation down is not that easy. We did not find a method to treat these data as an ensemble. This leads to the difficulty of integrating this pre-selection part into an FPGA. For now, this pre-selection part was implemented by matlab on PC.
 
In conclusion, our work has proven the possibility of set up a  cheap, practical, photon-free while stable QRNG. However, there still many aspects to be optimized. We hope our work could give some useful ideas in designing practical QRNGs.

\section{Acknowledgments}

We thank Pan Dong, Gao Xingyu, Liu yipu for their useful discussions on the structure of Si SPAD. We also thank Liu Xinyu and Ai fei for their assistance in designing and using software. Support from Beijing Advanced Innovation Center for Future Chip(ICFC) is gratefully acknowledged. We also appreciate financial support from National Nature Science Foundation of China(Grant No. 04130211).

\section{ Appendix: Detailed result of randomness tests}
The detailed data analysis by NIST test is obtained by the official program 'sts' version $2.1.2$, as shown in TABLE\ref{tab1} . And the detailed data analysis by Dihard test is shown as the following TABLE\ref{tab2}:

\begin{table}[h]
\centering
\begin{tabular}{lccc}\hline Statistical Test &P-value & Proportion & Assessment\\\hline Frequency & 0.761328 & 0.990119 & Success\\ BlockFrequency & 0.874053 & 0.990514 & Success \\
CumulativeSums & 0.897521 &0.989526 & Success\\ Runs & 0.979283 & 0.994664 & Success\\ LongestRun &0.345221&0.990316 & Success\\ Rank & 0.102931 &0.992688 & Success\\ FFT &0.764302 &0.989723 & Success\\NonOverlappingTemplate &0.223218 & 0.987945 & Success\\OverlappingTemplate & 0.298453 &0.990316 & Success\\ Universal & 0.496539 & 0.987945  & Success\\ ApproximateEntropy & 0.978821 & 0.992292 & Success\\RandomExcursions & 0.622942 & 0.993312 & Success\\RandomExcursionsVariant & 0.432522 & 0.986641 & Success \\Serial & 0.710092 & 0.989723& Success\\ LinearComplexity &0.447382 & 0.992095& Success \\\hline 
\end{tabular}

\caption{Result of NIST test for a $1.5GB$ final data, The minimum pass rate for each statistical test with the exception of the random excursion (variant) test is approximately = 1499 for a sample size = 5160 binary sequences. The minimum pass rate for the random excursion (variant) test is approximately = 517 for a sample size = 524 binary sequences. As the confidence parameter $\alpha = 0.01$, our data passed the NIST test. }
\label{tab1}
\end{table}

\begin{table}
\centering
\begin{tabular}{lcc}\hline StatisticalTest & P-value & Assessment\\\hline BirthdayTest & 0.498372 & Success\\ OverlappingPermutation & 0.387202 & Success \\RanksOf$31\times 31$matrices & 0.887632 & Success\\ RanksOf$32 \times 32$matrices & 0.342685 & Success\\ RanksOf$6 \times 8$matrices &0.912761 & Success\\ BitstreamTest &0.102923 & Success\\ OPSO &0.6521 & Success\\OQSO &0.8467 & Success\\DNA  & 0.7129 & Success\\ Count$1s$inTheStreamOfBytes& 0.273583 & Success\\ Count $1s$InTheSpecialBytes & 0.844794 & Success\\ParkingLotTest & 0.247862 & Success\\MinimumDistanceTest & 0.9221 & Success \\$3-D$ SpheresTest &0.776623 & Success\\ SqueezeTest& 0.386842 & Success \\OverlappingSumsTest 0.299432& Success\\ Runs & 0.756887& Success \\ Craps & 0.194702& Success\\\hline

\end{tabular}
\caption{Result of Diehard test for a $1.5GB$ final data, All of these indexes lies in$(0.000001,0.999999)$, our data passed the Diehard test. }
\label{tab2} 
\end{table}

\end{document}